      [1999/12/01 v1.4c Il Nuovo Cimento]
\documentclass{cimento}
\usepackage{graphicx}

\title{The Indirect Search for Dark Matter from the centre of the Galaxy with the Fermi LAT}

\author{A.~Morselli\from{ins:x}\ETC,
B.~Ca\~nadas\from{ins:x}  \atque
V.~Vitale\from{ins:x} \\  {\sl on behalf of the $Fermi$ LAT collaboration} 
     }
\instlist{\inst{ins:x} INFN sez. Tor Vergata, Roma, Italy}

\PACSes{
}
\begin{document}
\shorttitle{The Indirect Search for DM from the Galactic Center with the $Fermi$ LAT}

\maketitle

\begin{abstract}
Dark matter (DM) constitutes around a 25\% of the Universe, while baryons only a 4\%.
DM can be reasonably assumed to be made of particles,
and  many theories (Super-symmetry, Universal Extra
Dimensions, etc.) predict Weakly Interacting Massive Particles (WIMPs) as natural
DM  candidates at the weak scale.
Self-annihilation (or decay) of WIMPs might produce secondary $\gamma$ rays, via
 hadronization or as final state radiation.
Since its launch in the 2008, the Large Area Telescope on-board of the $Fermi$ $\gamma$-ray
Space Telescope has detected the largest amount of $\gamma$ rays to date, in the 20MeV 300GeV
energy range, allowing to perform a very sensitive indirect experimental 
search for DM (by means of high-energy $\gamma$-rays).
DM forms large gravitationally bounded structures, the halos, which
can host entire galaxies, such as the Milky Way.
The DM  distribution in the central part of the halos is not experimentally know, 
despite a very large density enhancement might be present.
As  secondary $\gamma$ rays production is very sensitive to WIMP density,
a very effective search can be performed  from the regions 
where the largest density is expected.
Therefore the information provided by the DM halo N-body simulations are crucial.
The largest $\gamma$-ray  signal from DM annihilation is expected from the centre of the Galaxy.
In the same region a large $\gamma$-ray background is produced by
bright discrete sources and the  cosmic-rays interacting with
 the interstellar gas and the photons fields.
 Here we report an update of the indirect search for
DM from the Galactic Center (GC).

\end{abstract}

\section{The Large Area Telescope on-board of $Fermi$}

The $Fermi$ $\gamma$-ray Space Telescope ($Fermi$)  was launched on June 11,
 2008 and began operations on August 11, 2008. The observatory carries
 two instruments for the study of $\gamma$-ray emission from astrophysical
 sources: the Large Area Telescope (LAT) and the $\gamma$-ray Burst Monitor (GBM).
LAT is the primary instrument and is a pair-conversion telescope.
It  is composed of a 4$\times$4 array of equal modules (towers) and surrounded
 by a segmented anti-coincidence detector (ACD). Each tower is made of
a precision silicon-strip tracker (36 layers arranged in 18 X-Y pairs alternating with W converter layers
 and a calorimeter.
The calorimeter is  a hodoscopic configuration of 8.6 radiation lengths (X$_{0}$)
 of CsI crystals that allows imaging of the shower development in the
 calorimeter and thereby corrections of the energy estimate for the
 shower leakage fluctuations out of the calorimeter.
The total thickness of the tracker and calorimeter is approximately 10 X$_{0}$ (1.5 X$_{0}$ for the tracker and 8.6X$_{0}$ for the calorimeter)
 at normal incidence. The ACD  covers the tracker array, and a programmable
 trigger and data acquisition system uses prompt signals available from
 the tracker, calorimeter and ACD to form a trigger that initiates readout
 of these three subsystems. The on-board trigger is optimized for rejecting
 events triggered by cosmic-ray background particles while maximizing the
 number of events triggered by $\gamma$-rays, which are transmitted to the ground
 for further processing.
The second instrument is  the $\gamma$-ray Burst Monitor (GBM), which  is a detector covering the
 8 keV-40 MeV energy range, devoted to the study of the $\gamma$-ray Bursts.
 GBM complements the LAT for observations of high-energy transients.
The GBM consists of two sets of six low-energy (8 keV to 1 MeV) NaI(Tl)
 detectors and a high-energy (0.2 to 40 MeV) BGO detector.
Detailed descriptions of the $Fermi$ observatory can be found in \cite{descr} and 
the LAT on-orbit calibration is reported in \cite{onorb}.

During the first two years, $Fermi$ operations have been mainly performed
 in the so-called "scanning" mode, with which the sky exposure is almost uniform. 
For autonomous repoints or for other targets of opportunity, the observatory
 can be inertially pointed

The Large Area Telescope has an effective area five times larger, a much
 better angular resolution, and a sensitivity more than 10 times better
 than its predecessor EGRET. The $Fermi$ LAT Collaboration has already detected 
 thousands of high energy $\gamma$-ray sources \cite{Fermi_catalog} and has carried out the 
study of  several scientific objectives during the first two years of operations.
These  studies  span many topics of astrophysics and fundamental physics.
Among the galactic sources Globular Clusters, Supernova Remnants, Binary Sources as well as  a large number of pulsars
have been detected and studied .
The Galactic diffuse emission has also been investigated.
Regarding the  extragalactic sources,  hundreds of
 Blazars and Active Galaxies, some Radio Galaxies, the Large Magellanic Cloud
and a couple of  Starburst Galaxies   have been  detected.
The nature of the extragalactic $\gamma$-ray background has been studied.
High energy $\gamma$ ray emission associated with  $\gamma$-Ray Bursts has
also been detected as well  as local $\gamma$-ray sources (Earth,  Sun and the Moon).

One of the major scientific objectives of the LAT is the indirect search for DM,
by means of the production of secondary $\gamma$-rays after the annihilation (or decay) of the DM particle candidates.
The search strategy, which was assessed with a detailed study \cite{prelaunch},
 comprises the study of targets with an expected relatively large 
$\gamma$-ray signal (such as the Galactic Center, which was previously studied with EGRET data \cite{dark}), or   with a very low foreseen conventional $\gamma$-ray emission \cite{sfer}, the search for annihilation lines \cite{lines} and also the search of possible 
anisotropies generated by the DM halo substructures \cite{ani}. The indirect DM searches with $\gamma$
rays are complemented with those performed with the detection of cosmic-ray electrons by the LAT \cite{Fermi_el}, \cite{Fermi_el2}.
In the next sections we provide an introduction on the indirect search for DM signal form the GC and an update of the results obtained so far by the $Fermi$ LAT Collaboration on this target.
\section{The Indirect Search for DM with High Energy $\gamma$ rays}
\begin{figure}[t]
\begin{minipage}[]{0.5\linewidth}
  \includegraphics[width=0.98\linewidth]{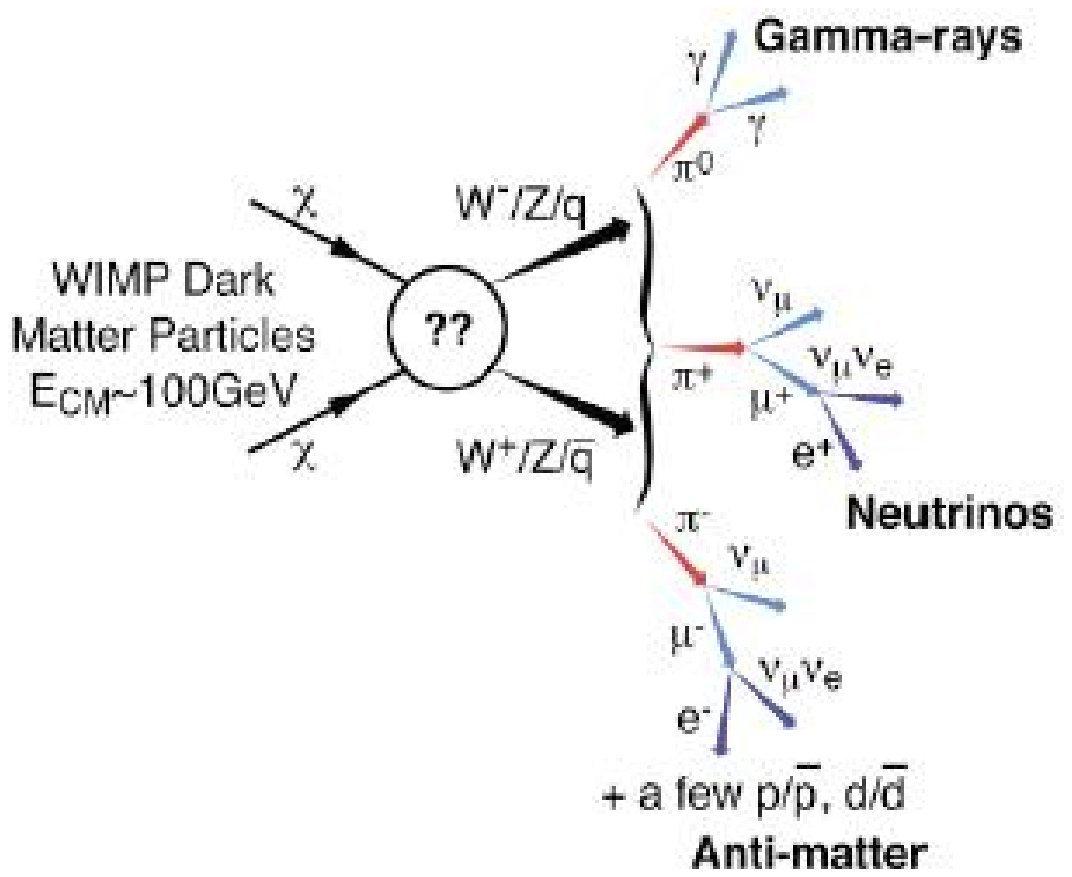}
\end{minipage}
\begin{minipage}[]{0.5\linewidth}
  \includegraphics[width=0.98\linewidth]{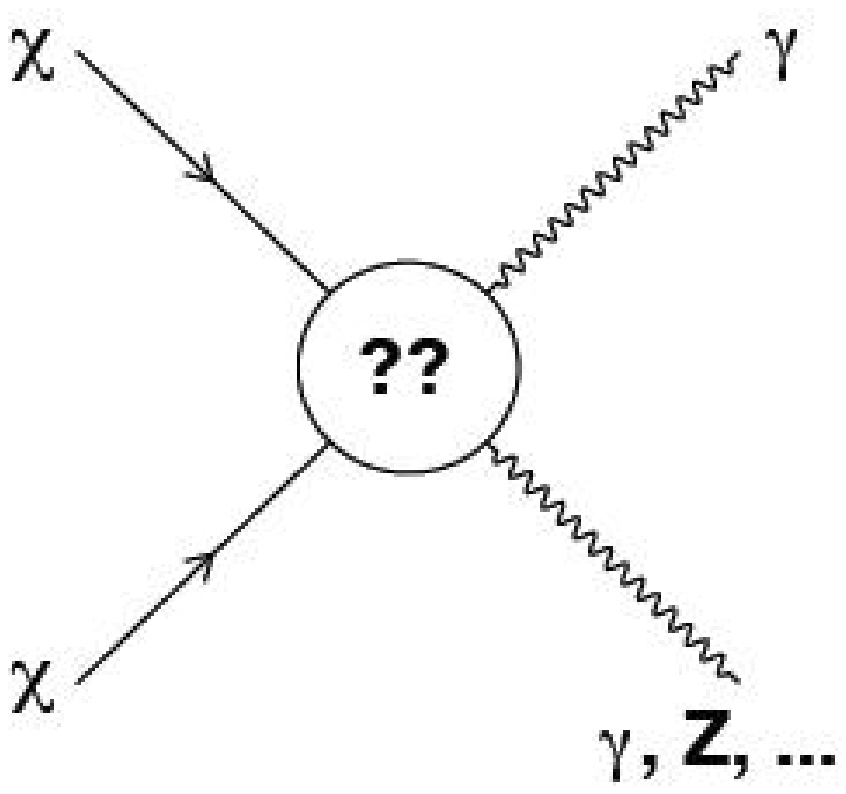}
\end{minipage}
\caption{Schematic view from reference \cite{prelaunch} of how gamma rays and
other secondary 
particles can be produced by WIMP annihilation. On the left figure the production of
a continuum 
spectrum is illustrated, whereas on the right the production of monochromatic lines
is shown.}
\label{fig:illustrative}
\end{figure}

\begin{figure}[ht]
\hskip -0.4 cm
  \includegraphics[width=0.45\linewidth,angle=-90]{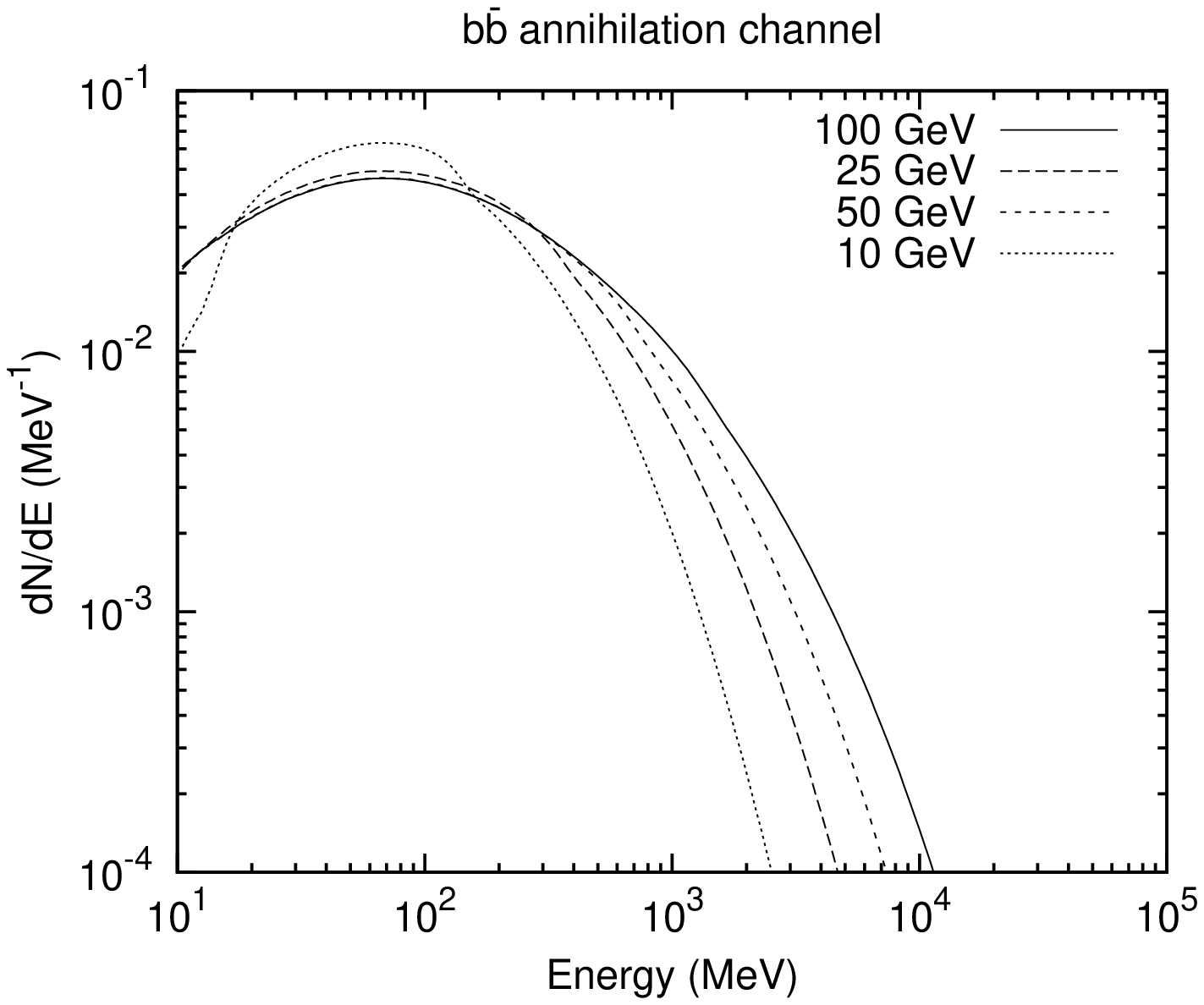}
\hskip -0.4 cm
  \includegraphics[width=0.45\linewidth,angle=-90]{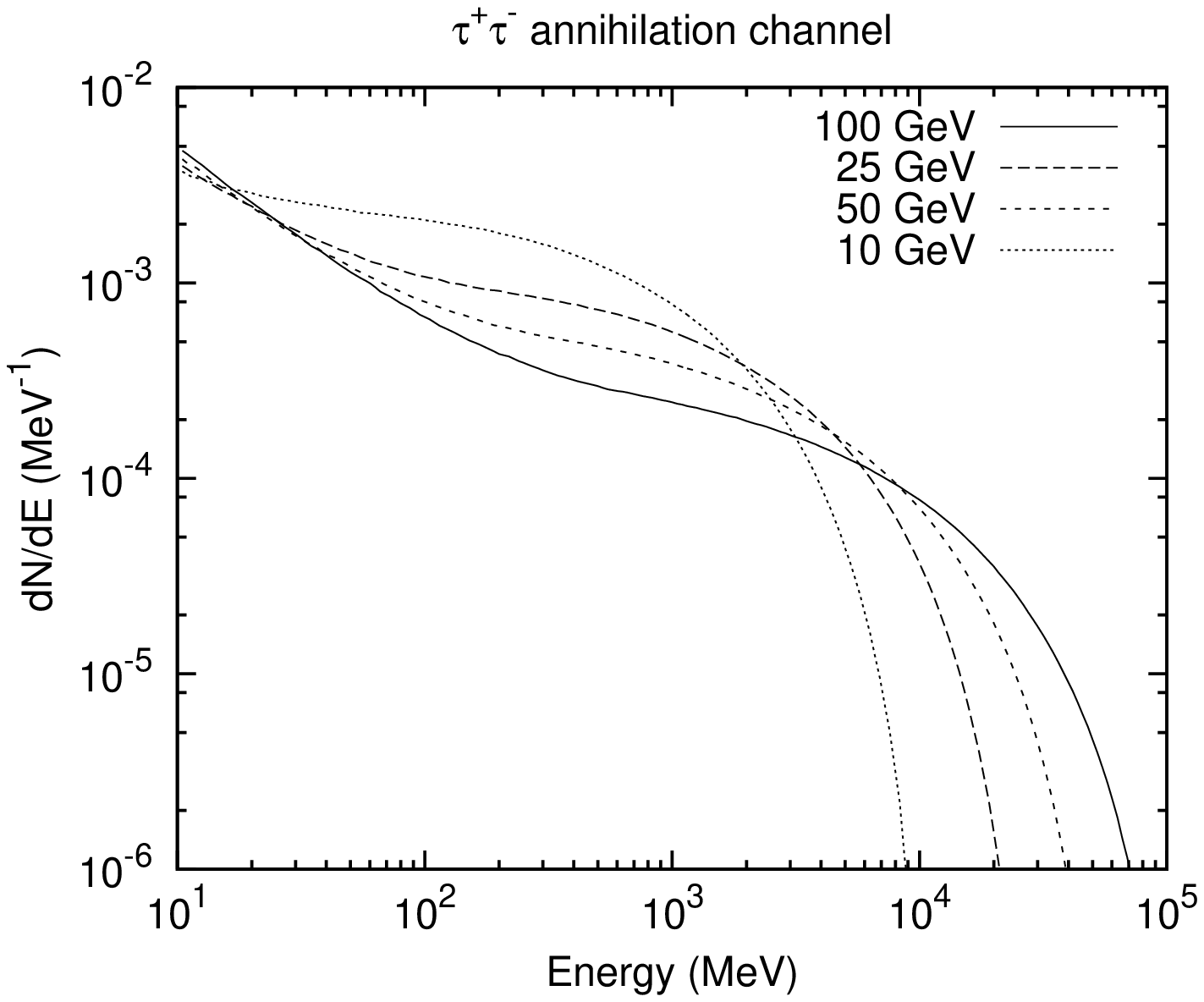}
\caption{Differential photon yield per DM annihilation.
In the left panel, we represent annihilation into $b\bar{b}$, and in the
right panel
annihilation into $\tau^+\tau^-$. These yields have been calculated using the
DarkSUSY code \cite{ds}.}
\label{fig:yields}
\vskip -0.4 cm
\end{figure}

In models in which DM is characterized by weak-scale interaction 
cross-sections (WIMP DM), DM particles can produce a 
gamma-ray signal by pair annihilation. 
As shown in figure \ref{fig:illustrative}, this signal can be either 
a monochromatic line if WIMPS annihilate directly into photons, or a continuum 
spectrum if they annihilate into a pair of intermediate particles 
($q\bar{q},W^+W^-,l^+l^-$) which would subsequently produce gamma rays. 
The former process is in general suppressed by around two orders of magnitude with 
respect to the latter.
The differential gamma ray flux from WIMP annihilation coming from a given direction 
in the sky $(\theta,\phi)$ and from a given solid angle $\Delta\Omega$ is

\begin{equation}
\label{eq:gflux}
\frac{d\Phi}{dE}=\sum_i \frac{1}{4\pi}\frac{<\sigma v>}{2m_{WIMP}^2}
br_i\frac{dN_i}{dE}
\int_{los,\Delta\Omega}{\rho^2(l,\theta,\phi)dld\Omega}\,,
\end{equation}

\noindent which can be factorized in two parts:

 
a) {Particle physics - related terms:} the thermally averaged annihilation 
cross-section $<\sigma v>$, the mass of the WIMP $m_{\mathrm{WIMP}}$ and the 
branching ratio $br_i$ for annihilation through the particular channel $i$ 
depend on the theory by which the WIMP is predicted. 
The photon yield per WIMP annihilation $dN_i/dE$, depends on the annihilation 
channel we are considering,
and on the mass of the WIMP. Figure \ref{fig:yields} shows the differential yields
for different WIMP masses and for two annihilation channels, namely the 
$ b\bar{b}$ and $\tau^+\tau^-$ channels, which are strongly motivated by 
supersymmetry. 


b) Astrophysics - related term:  the integral of the squared DM 
density along the line of sight depends uniquely on the DM distribution.
Usually it is expressed in terms of the factor $J(\theta,\phi,\Delta\Omega)$, 
which is defined as 

\begin{equation}
J(\theta,\phi,\Delta\Omega)=\frac{1}{8.5kpc}\left(\frac{1}{0.3 \mathrm{GeV}
cm^{-3}}\right)^2\int_{los,\Delta\Omega}{\rho^2(l,\theta\phi)dld\Omega}\,.
\end{equation}

%
%

The strategies and design of instruments for direct and indirect DM 
searches must lie on some theoretical assumptions. These assumptions are 
related to both the intrinsic nature of DM, and the interactions it 
is subjected to (particle physics - related terms in equation \ref{eq:gflux}), 
and also to how it is distributed in the Universe (astrophysics - related term). 
Regarding the latter issue, numerical cosmological simulations have proven 
to be a fundamental tool in the study of DM. 
Being weakly interacting, DM can be very well approximated by a 
collisionless fluid, its dynamics being driven only by the gravitational interaction. 
However, high non-linearity makes it impossible to study analytically its dynamics 
and the evolution of its distribution, and numerical simulations become essential.
These have shown that large amounts of DM accumulate 
in the center of galactic halos. The center of our galaxy is therefore
the closest and brightest
source of gamma-rays coming from annihilating DM and a very promising
target for DM searches. 
Over the last years, the development of new numerical algorithms and simulation codes
together with the rapid progress in computer technology have permitted to 
increase the resolution of cosmological simulations by several orders of magnitude 
and to produce some of the current state-of-the-art cosmological simulations, such as 
Via Lactea \cite{vialactea}, GHALO \cite{ghalo} or Aquarius \cite{aquarius}.
These confirm that the inner density profiles of halos are well fitted by functions
which are steeper and cuspier than the traditionally used Navarro Frenk and White 
profile \cite{nfw}.
However, moving yet further towards the center of the halos, predictions on the dark 
matter and total mass distribution require a realistic treatment of the baryons and 
their dynamical interactions with the DM. 
Indeed, the presence of baryons and the physical processes in which they are involved,
dramatically affect the distribution of DM in galaxies. In particular,
the DM density profiles can steepen through the adiabatic contraction due to 
dissipating baryons.
This adiabatic contraction can be implemented analytically within halo models
obtained in collisionless DM simulations, or one can attempt to include 
baryons in numerical simulations, as done, e.g. within the CLUES project \cite{clues}. 
However, there is still a large uncertainty on how baryons affect the distribution of 
DM in the center of our galaxy, which needs to be taken into account when
trying to disentangle a DM signal from the galactic center.


\begin{figure}[t]
\begin{center}
  \includegraphics[width=0.99\linewidth,height=7.2cm,angle=0]{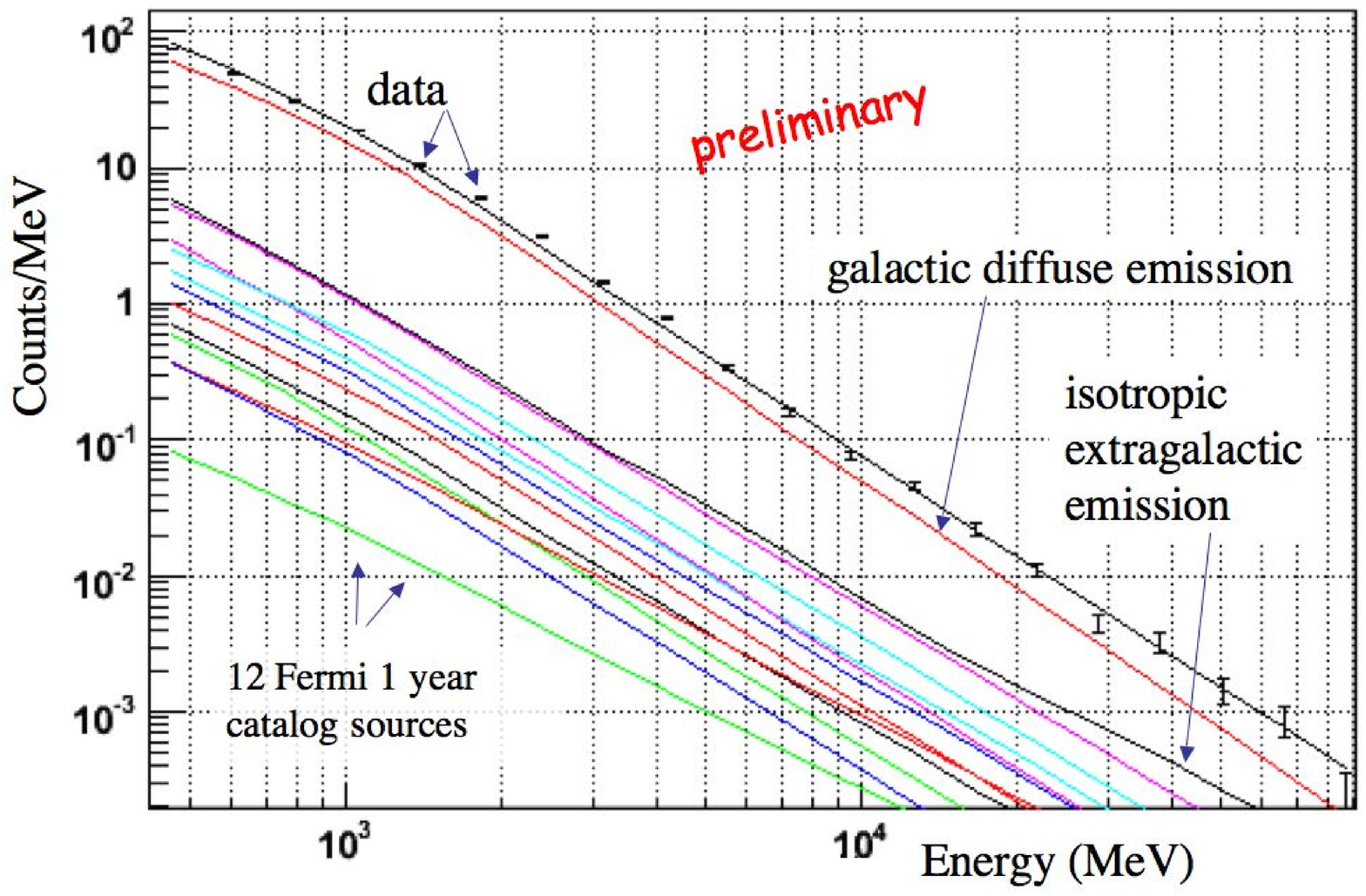}
\caption{Spectra from the likelihood analysis of the $Fermi$ LAT data (number of counts vs reconstructed energy) in a 7$^{\circ} \times $7$^{\circ}$ region around the Galactic Center (number of counts vs reconstructed energy)}
\label{fig:GC1}
\end{center}
%
%
%
%
\begin{center}
  \includegraphics[width=0.95\linewidth,height=4.2cm,angle=0]{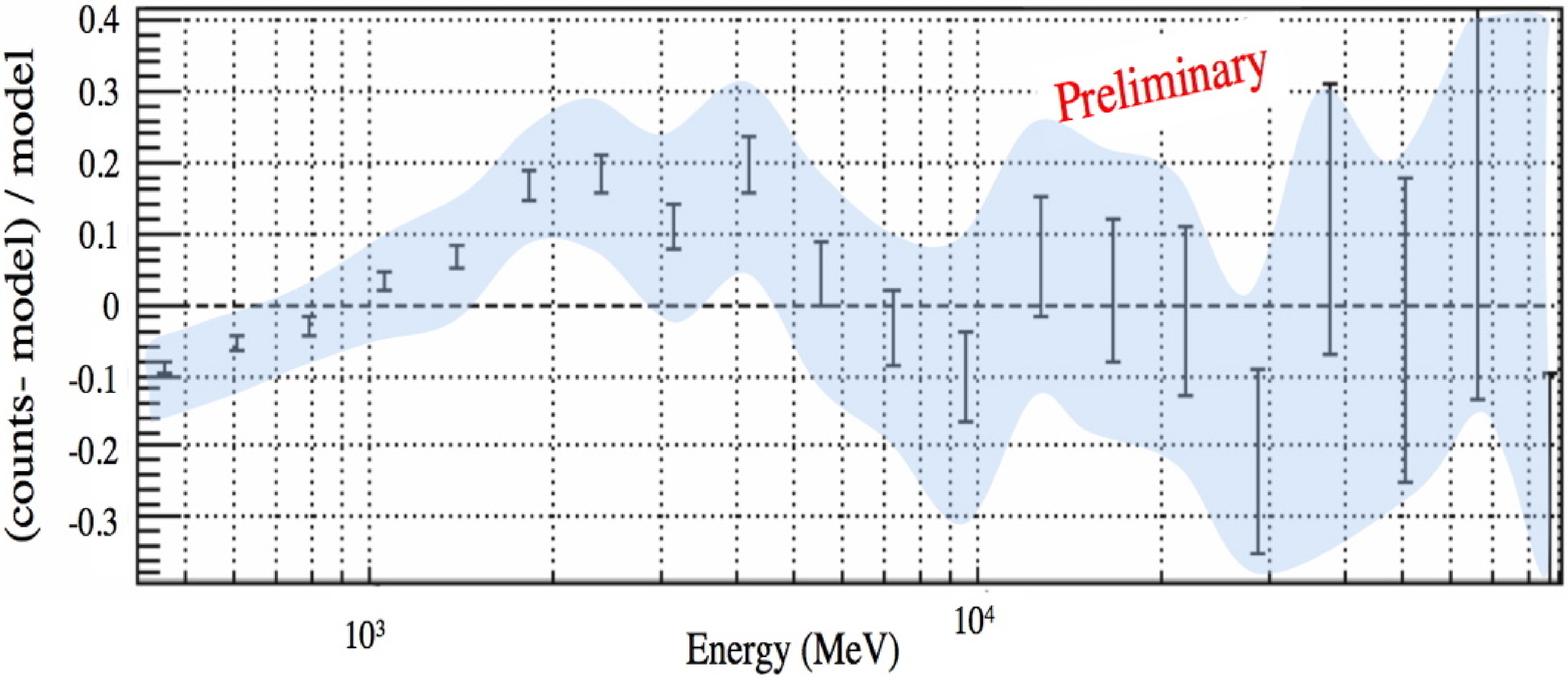}
\caption{Residuals ( (exp.data - model)/model) of the above likelihood analysis. The blue area  shows the systematic errors on the effective area.}
\label{fig:GC2}
\end{center}
\end{figure}

\section{Results on Indirect Search from  the Galactic Center}

A preliminary analysis of first 11 months of $Fermi$ LAT observations (August 2008 - July 2009) is reported.
A binned  likelihood analysis was performed with the  analysis software developed by the $Fermi$ LAT collaboration 
(gtlike, from the $Fermi$ analysis tools \cite{tools}). The P6$̣_{̣-}$v3 version of the Instrument Response Functions and 
event classification was used.  
For this  analysis a region of interest (RoI) of 7$^{\circ} \times $7$^{\circ}$ was considered in order to minimize
 the diffuse backgrounds contributions.
The RoI was centered at the Galactic Center position at RA = 266.46$^{\circ}$, Dec=-28.97$^{\circ}$.
The  events were selected to have an energy  between 400MeV and 100GeV, to be of the "diffuse" class
(high purity sample)  and to have converted in the \emph{front} part of the tracker. 
The selection conditions  provided us with events with very well reconstructed incoming direction.
Data have been binned into a 100$\times$100bins map for the subsequent likelihood analysis.
In order to perform maximum likelihood analysis of the data, a  model of the already known sources and the diffuse  background should be built.
The used model  is made of  11 sources from the $Fermi$ 1 year catalog  \cite{Fermi_catalog} which are located  within or very close to  the considered region being analyzed.
These sources have a point-like spatial model and a power-law spectrum in the form of a power-law.
The model also contains the diffuse $\gamma$-ray background which is made of two components:
 1) the  {\sl Galactic Diffuse $\gamma$-ray background} was modeled by means of the GALPROP code (model number 87XexphS) \cite{str} and \cite{str2};  2) the {\sl Isotropic Background} was modeled as an isotropic emission with a template spectrum and should account for both the Extragalactic $\gamma$-ray emission and residual charged particles in the data sample.
%
%
%
The results of the likelihood analysis are shown in  figures \ref{fig:GC1}  and \ref{fig:GC2}, for further details see \cite{symp}.
The diffuse $\gamma$-ray backgrounds and discrete sources, as we know them today, can account for the large majority of the 
detected $\gamma$-ray emission from the Galactic Center. Nevertheless a residual  emission is left, not accounted for by the above models.
Improved modelling of the Galactic 
diffuse model as well as the potential contribution from other 
astrophysical sources (for instance unresolved point sources) could 
provide a better description of the data. Analyses are underway to 
investigate these possibilities.

 

\acknowledgments

The $Fermi$ LAT Collaboration acknowledges support from a number of agencies and institutes for both development and the operation of the LAT as well as scientific data analysis. These include NASA and DOE in the United States, CEA/Irfu and IN2P3/CNRS in France, ASI and INFN in Italy, MEXT, KEK, and JAXA in Japan, and the K.~A.~Wallenberg Foundation, the Swedish Research Council and the National Space Board in Sweden. Additional support from INAF in Italy and CNES in France for science analysis during the operations phase is also gratefully acknowledged.

\end{document}